\documentclass{aa}  
\usepackage{graphicx}
\usepackage{txfonts}
\usepackage{amsmath}	% Advanced maths commands
\usepackage{amssymb}	% Extra maths symbols
\usepackage{xcolor}
\usepackage[colorlinks=true,citecolor=blue,linkcolor=blue]{hyperref}
\newcommand{\hess}{H.E.S.S.}
\newcommand{\src}{Westerlund~1}

\newcommand{\kms}{\,\mathrm{km}\,\mathrm{s}^{-1}}
\newcommand{\Msyr}{\,\mathrm{M}_\odot\,\mathrm{yr}^{-1}}
\newcommand{\ergs}{\,\mathrm{erg}\,\mathrm{s}^{-1}}

\newcommand{\eVqcm}{\,\mathrm{eV}\,\mathrm{cm}^{-3}}
\newcommand{\pqcm}{\,\mathrm{cm}^{-3}}
\newcommand{\Msun}{\,\mathrm{M}_\odot}

\newcommand{\Lw}{$L_\mathrm{w}$}

\defcitealias{Aharonian22}{HC22}

\begin{document}

   \title{Understanding the TeV $\gamma$-ray emission surrounding the young massive star cluster \src}
   \titlerunning{Understanding $\gamma$-ray emission around \src}
   \authorrunning{L. H\"arer et al.}

   %\subtitle{subtutle}

   \author{Lucia K. H\"arer\thanks{\email{lucia.haerer@mpi-hd.mpg.de}}, 
            Brian Reville, Jim Hinton, Lars Mohrmann, and Thibault Vieu}

   \institute{Max-Planck-Institut für Kernphysik, Saupfercheckweg 1, 69117 Heidelberg, Germany}

   \date{Received 14/11/2022; accepted 25/01/2023}

% \abstract{}{}{}{}{} 
% 5 {} token are mandatory

  \abstract
   {Young massive star clusters (YMCs) have come increasingly into the focus of discussions on the origin of galactic cosmic rays (CRs). The proposition of CR acceleration inside superbubbles (SBs) blown by the strong winds of these clusters avoids issues faced by the standard paradigm of acceleration at supernova remnant shocks.}
   {We provide an interpretation of the latest TeV $\gamma$-ray observations 
   of the region around the YMC \src\ taken with the High Energy Stereoscopic System (\hess) 
   in terms of diffusive shock acceleration at the cluster wind termination shock, taking into account the spectrum and morphology of the emission. As \src\ is a prototypical example of a YMC, such a study is relevant to the general question about the role of YMCs for the Galactic CR population.}
   {We generate model $\gamma$-ray spectra, characterise particle propagation inside the SB based on the advection, diffusion, and cooling timescales, and constrain key parameters of the system. We consider hadronic emission from proton-proton interaction and subsequent pion decay and leptonic emission from inverse Compton scattering on all relevant photon fields, including the CMB, diffuse and dust-scattered starlight, and the photon field of \src itself. The effect of the magnetic field on cooling and propagation is discussed. Klein-Nishina effects are found to be important in determining the spectral evolution of the electron population.}
   {A leptonic origin of the bulk of the observed $\gamma$-rays is preferable. The model is energetically plausible, consistent with the presence of a strong shock, and allows for the observed energy-independent morphology. The hadronic model faces two main issues: confinement of particles to the emission region and an unrealistic energy requirement.}
  % conclusions heading (optional)
   {}
   

   \maketitle
%
%-------------------------------------------------------------------

\section{Introduction}
\label{sec:intro}

The standard paradigm of cosmic ray (CR) acceleration in the Galaxy states that supernova remnants (SNR) are the dominant source class at ${\sim}\mathrm{GeV}\mbox{--}\mathrm{PeV}$. This scenario is faced with several long-standing problems \citep[for a review see][]{Gabici19}, for example the mis-match of models to the observed $^{22}$Ne/$^{20}$Ne ratio and the fact that acceleration to PeV energies is only conceivable under certain conditions, namely high shock velocities in dense environments. In addition, the $\gamma$-ray spectra of several SNR have been found to cut off early in the TeV-band, disfavouring recent PeV acceleration \citep[see][]{Funk15}. Superbubbles (SBs) forming around young star clusters and associations present an alternative scenario, as was recognised early on \citep{Cesarsky83}. Young star clusters typically contain hundreds of massive stars with powerful, supersonic winds \citep[see][]{PortegiesZwart10}. These winds blow a cavity in the remnant of the parent molecular cloud. If the cluster is sufficiently compact, a termination shock forms inside this cavity. Beyond the termination shock radius lies the hot, shocked interior of the SB, which is delimited at the outer edge by a contact discontinuity separating the wind from a thin shell of swept-up material \citep{Weaver77, MacLow88}. This picture applies after a brief initial phase when the shell becomes radiative. While observations have confirmed the early theory in broad strokes, there remains a systematic discrepancy in bubble size and temperature, which is known as the SB energy crisis \citep[see][Ch.~1.5]{Oey09, VieuPhD21}. Losses due to non-negligible interstellar pressure, a porous and thermally conducting shell, dust emission, and escaping non-thermal particles are suggested to reduce the energy available to inflate and heat up the SB. The analytic theory can be corrected by introducing an empirical pre-factor, $\xi_\mathrm{b}$, scaling the energy input by the wind. \citet{Vieu22} estimate $\xi_\mathrm{b} \approx 22\%$ from observations, a value that is in the range predicted by current simulations \citep[e.g.][]{Gupta18b}.

Several models explore the SB scenario for galactic CRs \citep[e.g.][]{Bykov01, Ferrand10, Morlino21, Vieu22}, discussing multiple sites with favourable conditions for particle acceleration: the cluster wind termination shock, the turbulent bubble interior, and the cluster itself. The termination shock can reach high Mach numbers and potentially accelerate particles to PeV energies \citep[see recent works by][]{Morlino21, Vieu22b} due to its size \citep[\({\sim}10\)s of pc,][]{Weaver77} and the high velocities expected for cluster winds \citep[\({\sim}2000\mbox{--}3000\kms\),][]{Stevens03}. In the turbulent bubble interior, particles could then be stochastically reaccelerated by a Fermi-II-type process \citep{Klepach00, Bykov01}. Particles could also be accelerated by supernovae (SNe) exploding in the cluster and expanding inside the low-density wind, therefore reaching higher shock velocities, or at colliding winds inside the cluster \citep[see, e.g. discussion in][]{Vieu22}.  

In the last decade, evidence for particle acceleration in star cluster environments has accumulated through $\gamma$-ray observations \citep[][]{Ackermann11, Abramowski12, HESS15, Yang18, Aharonian19, Abeysekara21, Mestre21}. In this work, we examine the TeV $\gamma$-ray observations of the region around \object{Westerlund~1} taken with the High Energy Stereoscopic System (\hess) and reported in \citet{Aharonian22} (henceforth \citetalias{Aharonian22}). \src\ is a young \citep[4--5\,Myr,][]{Beasor21}, massive ($3\mbox{--}5\times 10^{4}\,\Msun$, \citealt{Brandner08}, see also \citealt{PortegiesZwart10} and \citealt{Lim13}), compact (half-mass radius ${\sim}1$\,pc) star cluster, located at a distance of ${\sim}4\,$kpc from Earth and ${\sim}4.6$\,kpc from the Galactic Center (GC) \citep{Kothes07, Davies19, Negueruela22}. The cluster contains a large collection of young massive stars, including 24 Wolf-Rayet (WR) stars\footnote{To be specific, 16 WN and 8 WC stars. WN and WC are WR types for which nitrogen and carbon dominate the spectrum, respectively.}, one Luminous Blue Variable (LBV), 10 Yellow Hypergiants (YHGs) and Red Supergiants (RSGs), and several bright OB supergiants \citep{Clark20}. Extended TeV $\gamma$-ray emission from the vicinity of the cluster was first reported by the \hess\ collaboration in \citet{Abramowski12} and designated \object{HESS~J1646$-$458}. A \textit{Fermi} analysis by \citet{Ohm13} revealed a GeV source, slightly off-set from the TeV emission and the cluster. \citet{Muno06} detected diffuse, non-thermal X-rays extending at least 5' outwards from the cluster. The 2022 \hess\ results \citepalias{Aharonian22} reveal an emission region ${\sim}1.5^\circ$ in diameter which is centred just slightly off the cluster position (see Fig.~\ref{fig:hess_w1}). The emission has a ring-like, energy-independent morphology, and a $\gamma$-ray spectrum with spectral index ${\sim}2.4$ up to ${\sim}80\,\mathrm{TeV}$, which is constant across the source within the range of uncertainty. The total $\gamma$-ray luminosity is $9\times 10^{34}\ergs$ between $0.37$ and 100\,TeV, adopting a distance of $3.9$\,kpc. \citetalias{Aharonian22} conclude that particle acceleration in the \src\ SB or inside the cluster itself is the most promising interpretation of the results. This idea is supported by work based on the first \hess\ publication on the source \citep[e.g.][]{Aharonian19}. The region harbours other source candidates, but they do not provide sufficient power or cannot account for the extent of the emission \citepalias{Aharonian22}. Here, we investigate the scenario of particle acceleration at the cluster wind termination shock in the light of the 2022 \hess\ results. In Sect.~\ref{sec:acc-in-sb}, we constrain properties of the SB and the cluster wind and outline basic requirements for shock acceleration. In Sect.~\ref{sec:morph} we characterise the morphology expected from proton and electron cooling and transport. Section~\ref{sec:spectrum} discusses model spectra. Section~\ref{sec:concl} then combines the findings to a full picture. 

\section{Characterisation of the \src\ region} 
\label{sec:acc-in-sb}
In this section, we characterise the \src\ region and estimate key parameters. We discuss the substantial uncertainty afflicting many parameters and select fiducial values for the following analysis, which are summarised in Table~\ref{tab:pars}.

\renewcommand{\arraystretch}{1.2}
\begin{table}
    \centering
    \caption{Input parameters used in the default scenario. For the dependent parameters, see the text and the equations.}
    \begin{tabular}{l c l}
    \hline
    \hline 
    Par. & Value & Description \\
    \hline
    $t_\mathrm{sys}$  & 4\,Myr & cluster age  \\
    $d$  & 3.9\,kpc & cluster distance \\
    $L_\mathrm{w}$  & $10^{39}\ergs$ & cluster wind power \\
    $\dot{M}$  & $5\times 10^{-4}\Msyr$ & mass-loss rate \\
    $B$  & $2\,\mu$G & magnetic field (emission region) \\
    $B_\mathrm{acc}$  & $2\,\mu$G & magnetic field (acceleration region) \\
    $n_\mathrm{ext}$  & $100\pqcm$ & external density, outside SB \\    
    $L_\mathrm{bol}$  & $10^{41}\ergs$ & cluster bolometric luminosity\\
    $T_\mathrm{eff}$  & $40{,}000\,$K & cluster effective temperature  \\
    $\xi_\mathrm{b}$  & $22\%$ & empirical $L_\mathrm{w}$ scaling factor \\
    \end{tabular}
    \label{tab:pars}
\end{table}

\subsection{The superbubble}
\label{sec:bubble}
The size, termination shock position, and interior density of the SB are largely determined by cluster wind characteristics, the density of the environment, which we term external density, and the age of the cluster. A key parameter is the cluster wind power, which is defined as $L_\mathrm{w} = 0.5 \dot{M} v_\mathrm{w}^{2}$, with the mass-loss rate, $\dot{M}$, and the wind speed, $v_\mathrm{w}$. $L_\mathrm{w}$ is set by the sum of the contributions from stellar winds and SNe, which we discuss in turn. The WR wind power can serve as a lower bound on the former, neglecting winds from other star classes. At a rather low mass of 10--15$\Msun$, a WR star is expected to have an average wind power of $L_\mathrm{w} = 10^{37}\ergs$ \citep{Seo18}. With its 24 WR stars, \src\ therefore has $L_\mathrm{w} \gtrsim 2.4\times 10^{38}\ergs$. For a more accurate estimate, we calculate the time-dependent wind power of a toy-model cluster. To accommodate uncertainty we vary the key parameters around a range of plausible values. We first populate the cluster according to an initial mass function (IMF), $\xi(m)\propto m^{-\Gamma}$, assuming a cluster mass of $M = 3\mbox{--}5\times 10^{4}\Msun$ and lower and upper bounds for stellar masses of 0.4\,$\Msun$ and 120\,$\Msun$. As \src\ has a top-heavy IMF \citep{Lim13}, $\Gamma$ is set to $1.8\mbox{--}2.0$. Then, we let the cluster evolve by computing the mass-loss in each time-step and updating stellar masses accordingly. In addition, stars that surpass their lifetime according to \citet{Limongi06} are removed or enter a WR phase if $M_\mathrm{star} > 20\Msun$. Wind power, mass-loss, and WR phase lifetime are taken from \citet{Seo18}. We adjust the WR population in the model cluster to that of \src\ at the time the wind power is evaluated. The resulting cluster wind power at $4\mbox{--}5\,\mathrm{Myr}$ is $1.6\mbox{--}2.8\times 10^{39}\ergs$. The same approach can be used to estimate the average power hitherto ejected by SNe, which yields $3.7\times 10^{38}\mbox{--}1.5\times 10^{39}\ergs$, assuming $10^{51}\,$erg per SN. Taking the WR power discussed in the beginning of this section as a lower limit for the wind power, the estimate for the total power is ${\sim}0.5\mbox{--}4\times 10^{39}\ergs$. The $\gamma$-ray luminosity cited in Sect.~\ref{sec:intro} amounts to ${\sim}0.002\mbox{--}0.02\%$ of this estimate, a broadly plausible value for the combined efficiencies of particle acceleration and $\gamma$-ray emission. For the following analysis, we conservatively select $L_\mathrm{w} = 10^{39}\ergs$ as a fiducial value.

For the mass-loss, we again obtain a lower estimate from \citet{Seo18}. A WR star of $10\Msun$ loses $2\times 10^{-5}\Msyr$, which results in a cluster mass-loss of $\dot{M} \sim 5\times 10^{-4}\Msyr$. As WR have mass-loss rates greatly exceeding those of main sequence stars, we assume they dominate the total mass-loss and take $\dot{M}_\mathrm{cl} = 5\mbox{--}10\times 10^{-4}\Msyr$. For the cluster wind velocity, we obtain
\begin{equation}
    v_\mathrm{w} \sim 2500 \left( \frac{L_\mathrm{w}}{10^{39}\ergs} \right)^{0.5} \left( \frac{\dot{M}}{5\times10^{-4}\Msyr} \right)^{-0.5} \kms ,
\end{equation}
which means that $v_\mathrm{w}$ takes values of ${\sim}1300\mbox{--}5000\kms$ for the ranges of \Lw\ and $\dot{M}$ given above. Note that the lower and upper bound represent quite extreme edge cases. Typical $v_\mathrm{w}$ are 2000--3000$\kms$ \citep[e.g.][]{Stevens03}.

In addition to the characteristics of the wind, the external density, $n_\mathrm{ext}$, is needed to determine the extent of the SB. Neutral gas tracers reveal an average density of 10.5$\pqcm$ for H$_2$, as traced by CO and 3.2$\pqcm$ for HI, from 21\,cm observations, assuming a distance of 3.9\,kpc \citepalias{Aharonian22}. However, the distribution of material is not homogeneous and can be as high as 190$\pqcm$, as measured in one CO cloud. In addition, a significant fraction of the material is likely to be ionised due to the young cluster. This material and also dust are not seen by the tracers discussed above. We therefore take $n_\mathrm{ext} = 100\,\pqcm$ as a fiducial value for the following analysis and $10\,\pqcm$ as a lower bound. This range gives a density in the bubble interior of $n_\mathrm{int} = 0.02\mbox{--}0.10\pqcm$ \citep{MacLow88}. In the default scenario  (Table~\ref{tab:pars}), $n_\mathrm{int} = 0.078\pqcm$. Note that $n_\mathrm{int}$ weakly depends on $\xi_\mathrm{b}$ ($n_\mathrm{int} \propto \xi_\mathrm{b}^{6/35}$), a scaling factor which encompasses losses due to the SB energy crisis (see Sect.~\ref{sec:intro}).

\begin{figure}
    \centering
    \includegraphics[height=0.25\textheight]{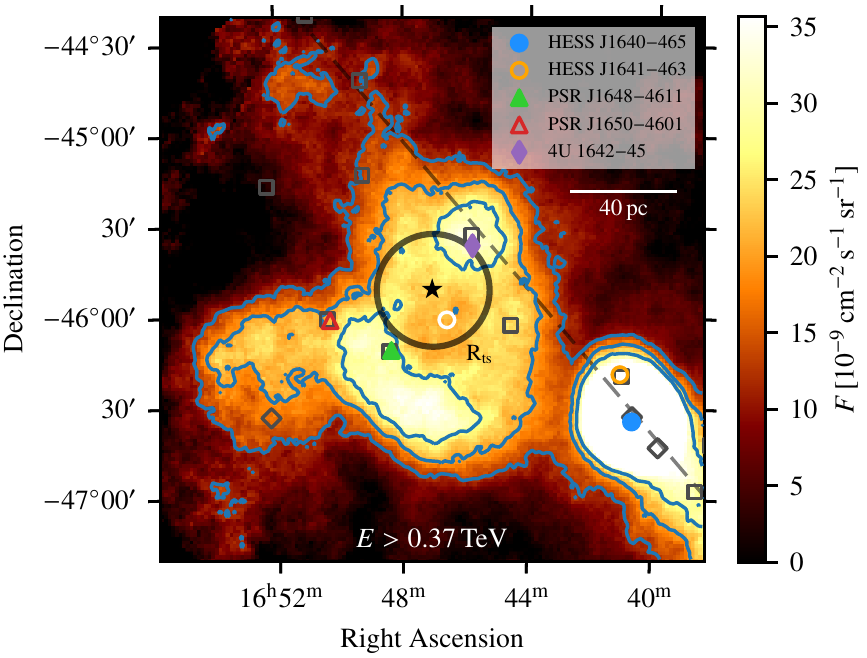}
    \caption{Map of HESS~J1646$-$458 taken from \citetalias{Aharonian22}, overlayed with the termination shock position predicted for the parameters indicated in Eq.~\ref{eq:rts}. The star marks the position of \src. The dashed grey line traces the galactic plane.}
    \label{fig:hess_w1}
\end{figure}

The estimates for wind power, mass-loss, and external density let us calculate the position of the forward shock,  $R_\text{b}$, and the termination shock, $R_\text{ts}$, which are 
\begin{multline}
    \label{eq:rts}
    R_\text{ts} = 20.4 \left( \frac{\xi_\mathrm{b}}{0.22} \right)^{-1/5} \left( \frac{\dot{M}}{5\times10^{-4}\Msyr} \right)^{3/10} \left( \frac{n_\mathrm{ext}}{100\,\mathrm{cm}^{-3}} \right)^{-3/10} \\ \times \left( \frac{v_\text{w}}{2500\kms} \right)^{1/10} \left( \frac{t_\text{sys}}{4\,\mathrm{Myr}} \right)^{2/5} \, \mathrm{pc} \, , 
\end{multline}    
\begin{multline}    
    \label{eq:rfs}
    R_\text{b} = 74.2 \left( \frac{\xi_\mathrm{b}L_\text{w}}{0.22\times 10^{39}\ergs} \right)^{1/5} \left( \frac{n_\mathrm{ext}}{100\,\mathrm{cm}^{-3}} \right)^{-1/5}\\ \times \left( \frac{t_\text{sys}}{4\,\mathrm{Myr}} \right)^{3/5} \mathrm{pc} \, ,
\end{multline}
according to analytic SB theory \citep{Weaver77}, where $L_\mathrm{w}$ was rescaled by $\xi_\mathrm{b}$ \citep[see][]{Vieu22}. At 3.9\,kpc, $R_\text{ts} = 20\,$pc corresponds to an angular radius of ${\sim}0.3^\circ$, which coincides well with the inner boundary of the $\gamma$-ray ring (see \citetalias{Aharonian22} and qualitatively Fig.~\ref{fig:hess_w1}). Allowing for the uncertainty in the parameters discussed earlier yields $R_\text{ts} \sim 20\mbox{--}60\,$pc and $R_\mathrm{b} \sim 60\mbox{--}180\,$pc. $R_\text{ts} = 60\,$pc would place most of the \hess\ emission upstream of the termination shock, which is implausible in a scenario where acceleration mainly occurs at the termination shock, leading us to disfavour the combination of high wind power and low external density. 

Finally, note that the \hess\ emission is slightly asymmetric and off-set from the cluster. Clumps and density inhomogeneities in the surrounding medium are a plausible origin of these features, especially as the elongation is perpendicular to the galactic plane (see Fig.~\ref{fig:hess_w1}). Asymmetry is a common feature of SBs, due to the inherent clumpiness of molecular clouds \citep[see][]{Chu08}. 

\subsection{The magnetic field in the acceleration region}

For particle acceleration at the termination shock to power the HESS J1646$-$458 emission, the shock must, in addition to being super-Alfv\'enic, have a magnetic field in the acceleration region, $B_\mathrm{acc}$, that allows for the observed maximum energy. The \hess\ spectrum lacks a clear cut-off and extends up to ${\sim} 80\,$TeV, from which we infer the parent electron and proton energies to be $E_\mathrm{max}^\mathrm{e} \gtrsim 100\,$TeV and $E_\mathrm{max}^\mathrm{p} \gtrsim\,480\,$TeV. With these values, Hillas' limit \citep{Hillas84} constrains the upstream $B_\mathrm{acc}$ to
\begin{equation}
    B_\mathrm{acc} > \frac{E_\mathrm{max}c}{ev_\mathrm{w}R_\mathrm{ts}} = 0.7\mbox{--}3.1 \left( \frac{v_\mathrm{w}}{2500\kms}\right)^{-1} \left( \frac{R_\mathrm{ts}}{20\,\mathrm{pc}}\right)^{-1} \,\mu\mathrm{G} \, ,
    \label{eq:hillas}
\end{equation}
for electrons and protons, respectively. In addition, efficient particle acceleration requires $M_\mathrm{A} \gg 1$, where the Alfv\'enic Mach number, $M_\mathrm{A}$, is the ratio of the wind velocity to the Alfv\'en speed, $v_\mathrm{A} = B_\mathrm{acc} /\sqrt{4 \pi \rho(R_{\rm ts})}$. We relate the gas density $\rho$ to the mass-loss, $4 \pi \rho(R) R^2 v_\mathrm{w} = \dot{M}$, where $R$ is the radial distance from the cluster, to obtain
\begin{equation}
    M_\mathrm{A} = \frac{v_\mathrm{w}}{ v_\mathrm{A}} = \frac{\sqrt{\dot{M}v_\mathrm{w}}}{B_\mathrm{acc} R_{\rm ts}} \, .
    \label{eq:mach}
\end{equation}
Upstream of the shock, we require
\begin{multline}
    B_\mathrm{acc} < 4.5 \left( \frac{R_\mathrm{ts}}{20\,\mathrm{pc}} \right)^{-1} \left( \frac{\dot{M}}{5\times 10^{-4}\Msyr} \right)^{0.5}\\ \times 
     \left( \frac{M_{\rm A}}{10}\right)^{-1}  \left( \frac{v_\mathrm{w}}{2500\,\kms} \right)^{0.5} \, \mu\mathrm{G} \, ,
    \label{eq:Bmax}
\end{multline}
where $M_{\rm A} =10$ was chosen as a reference value for a strong shock. Jointly with Eq.~\ref{eq:hillas}, Eq.~\ref{eq:Bmax} places a strong constraint on $B_\mathrm{acc}$ for the standard parameters that are given, especially for protons as they require a higher $E_\mathrm{max}$. The parameter space is especially narrow if the wind is slow. We select $B_\mathrm{acc} = 2\,\mu$G for the following analysis.

\subsection{The cluster photon field}
\label{sec:clphotfield}

\src\ contains a large sample of luminous stars, as discussed in the introduction. To obtain a lower bound for the bolometric luminosity of the cluster, we sum those of the brightest stars, for which measurements are available\footnote{The sample includes 18 WR stars \citep[16 WN and 2 WC,][]{Crowther06}, 6 YHGs \citep{Clark05}, and the 26 OB super-giants which have an absolute magnitude larger than $-6$ in \citet{Negueruela10}.}, which yields $L_\mathrm{bol} > 5 \times 10^{40}\ergs$. The sample only contains a subset of the brightest stars, we therefore conservatively adopt an estimate $L_\mathrm{bol} \sim 10^{41}\ergs$. Comparing this value to our estimate for the wind power, we note that $L_\mathrm{bol} \approx 100  L_\mathrm{w}$ presents a good approximation. The energy density of the cluster photons is $U_\mathrm{cl} = L_\mathrm{bol} (4\pi r^2 c)^{-1} \sim 42 (L_\mathrm{bol}/10^{41}\ergs) (R_\mathrm{ts}/20\,\mathrm{pc})^{-2} \eVqcm$. In the following, $U_\mathrm{cl}$ is always evaluated at the termination shock and therefore serves as an upper bound to the energy density inside the SB, at $R > R_\mathrm{ts}$. We model the emission from the cluster as thermal emission. In our sample, the YHGs are as luminous as the WR stars and the OB super-giants combined, which makes it difficult to draw a clear conclusion on whether hot or cool stars dominate the emission. We therefore investigate a broad range of effective temperatures for the cluster, $T_\mathrm{eff} = 10{,}000\mbox{--}50{,}000\,$K. 
The threshold electron energy above which Klein-Nishina (KN) suppression is important at these temperatures is \(E_{\rm KN} = 75 (T/10^4\,\mathrm{K})^{-1}\)\,GeV 
\cite[e.g.][]{BG70}. Thus, despite its large energy density, the cluster photon field does not result in a cooling time of the order of a few kyr, as would be expected for TeV particles in the Thomson regime. 

\section{Morphology} 
\label{sec:morph}

Two notable features of the source are its ring-like shape and energy-independent morphology. In the previous section, we showed that the position of the termination shock correlates with the size of the $\gamma$-ray ring. The morphology and extent of the emission region is determined by the transport and cooling of particles in the SB interior, which we investigate in this section. We also discuss the requirements to obtain an energy independent morphology.

\subsection{Cooling and transport -- Protons}
\label{sec:morph_protons}
The cooling time for protons through proton-proton (PP) interactions in the interior of the bubble can be written as \citep[e.g.][]{Aharonian04}
\begin{equation}
    t_\mathrm{cool} = (0.5 n_\mathrm{int} \sigma_\mathrm{pp} c)^{-1} \approx 5.2 \times 10^{8} x \left(\frac{n_\mathrm{int}}{0.1\pqcm}\right)^{-1} \,\mathrm{yr}  \, ,
    \label{eq:tcool_p}
\end{equation}
where we use the cross-section, $\sigma_\mathrm{pp}$, of \citet{Kafexhiu14}, normalised such that the scaling factor $x=1$ at 10\,TeV. The value of $x$ varies by less than $50\%$ in the TeV--PeV range.  Clearly, $t_\mathrm{cool} \gg t_\mathrm{sys}$, as $n_\mathrm{int} \leq 0.1\pqcm$ (see Sect.~\ref{sec:acc-in-sb}). This result has two important consequences: firstly, the efficiency of $\gamma$-ray production via PP interactions inside SBs is low, of the order $t_\mathrm{sys}/t_\mathrm{cool}$, which increases the power requirement. Secondly, the size of the emission region scales with the age of the system. We discuss the consequences of advection and diffusion in turn. A common result of SB models is that the density is approximately constant inside the SB. Mass continuity requires $v\rho R^2 = \mathrm{const.}$, such that $\rho(R)= \mathrm{const.} \Rightarrow v = \dot{R}\propto R^{-2}$. The flow velocity immediately downstream of the termination shock is $ v(R_{\rm ts})\approx v_\mathrm{w}/4$ in the strong shock limit. The time taken to advect from the shock to the inferred gamma-ray source radius $R_{\rm source}\approx 60$\,pc is obtained by solving the above differential equation:
\begin{multline}
    t_\mathrm{adv} = 271 \left( \frac{v_\mathrm{w}}{2500\kms} \right)^{-1} \left( \frac{R_\mathrm{ts}}{20\,\mathrm{pc}} \right)^{-2} \\ \times \left[ \left( \frac{R_\mathrm{source}}{60\,\mathrm{pc}} \right)^3 - \left( \frac{R_\mathrm{ts}}{20\,\mathrm{pc}} \right)^3 \right] \,\mathrm{kyr} \, ,
    \label{eq:tadv}
\end{multline}
which is much shorter than the age of the system. In fact, protons are expected to fill the interior of the SB completely. This finding is inconsistent with the size of HESS~J1646$-$458, unless the wind is exceptionally weak, such that $R_\mathrm{b} = R_\mathrm{source} \approx 60\,$pc. In addition, the Bohm limit places a lower bound on the diffusive transport, assuming isotropy. We parameterise the diffusion coefficient as  
\begin{equation}
    D = \frac{1}{3} h(E) r_\mathrm{g} c \, .
    \label{eq:bohmdiff}
\end{equation}
where we introduce the Hall parameter, $h(E)$, as the ratio of a particle's mean free path to its gyroradius, $h = \lambda/ r_\mathrm{g}\geq 1$, with $r_\mathrm{g} = E(eB)^{-1}$. The diffusion length is $\ell_\mathrm{p} = \sqrt{4Dt_\mathrm{sys}}$, which, in the Bohm limit ($h=1$), can be written as
\begin{equation}
    \ell_\mathrm{p} = 94 \left( \frac{E}{10\,\mathrm{TeV}} \right)^{0.5} \left( \frac{t_\mathrm{sys}}{4\,\mathrm{Myr}} \right)^{0.5} \left( \frac{B}{2\,\mu\mathrm{G}} \right)^{-0.5} \mathrm{pc} \, ,
    \label{eq:ellp}
\end{equation}
where $B$ is the magnetic field in the emission region. From Eq.~\ref{eq:ellp} and $R_\mathrm{ts} \gtrsim 20\,$pc, it follows that protons could fill the SB interior by diffusion alone, even for high $B$ such as $10\,\mu$G. Considering that diffusion and advection are both at play and that particles can diffuse across the contact discontinuity, which extends the emission region beyond $R_\mathrm{b}$, these findings effectively rule out that the observed morphology is dominated by protons. The effect of radial diffusion would have to be essentially reduced to zero, in addition to the presence of a weak wind causing the bubble to be small. Furthermore, such a scenario would raise the question why the emission region does not coincide with the expected location of the shell, despite its large density (${\sim} 10^{3}\,n_\mathrm{int}$, see \citealt{Gupta18}). The scenario that the observed emission is shell emission, would require $R_\mathrm{b} \approx 20\,$pc, which is inconsistent with SB theory and observations (see Sect.~\ref{sec:bubble} and \citealt{Chu08}).

\subsection{Cooling and transport -- Electrons}
\label{sec:morph_e}

\begin{figure*}
    \centering
    \includegraphics[height=0.25\textheight]{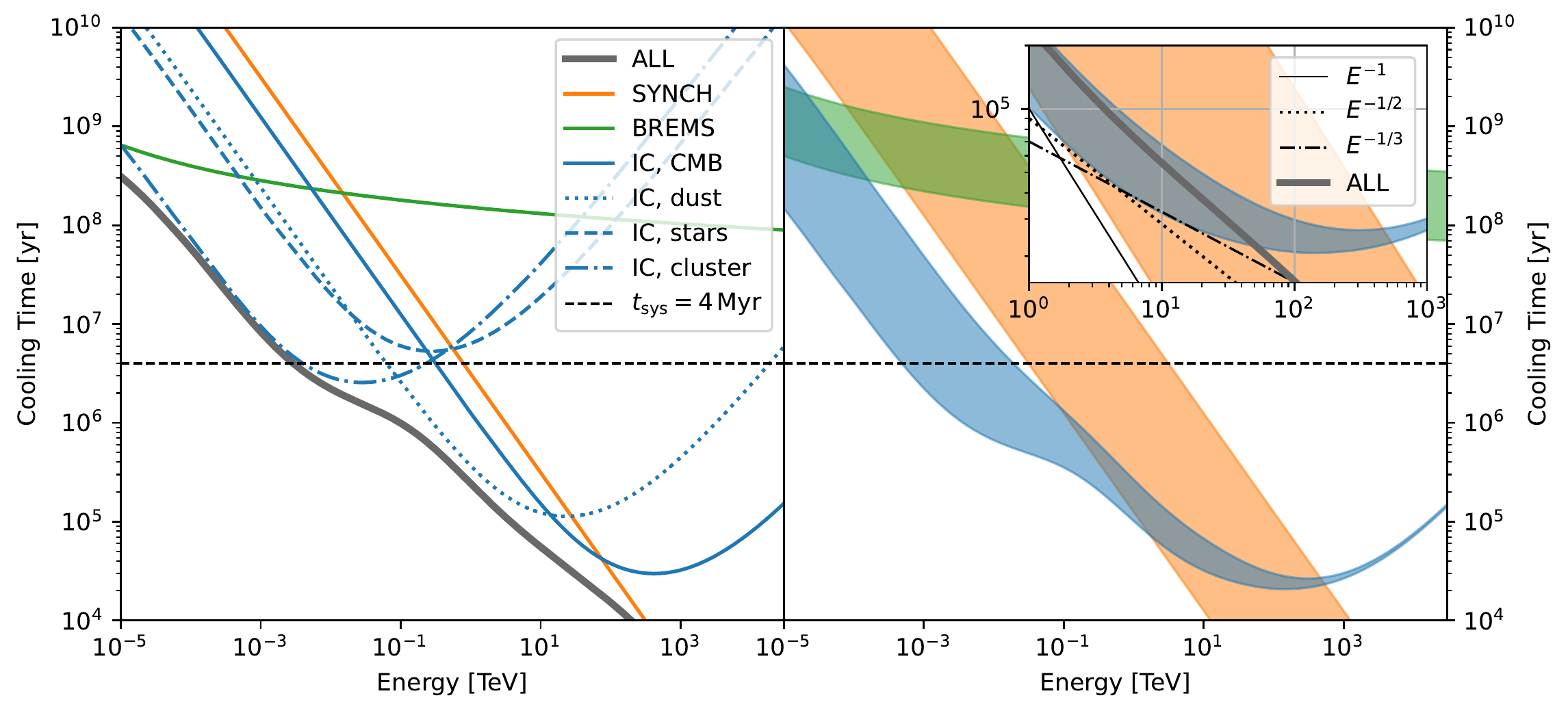}
    \caption{\textit{Left:} electron cooling times for the default case (see Table~\ref{tab:pars}). The broad grey line is the sum of all the components shown in colour. For a description of the photon fields for IC scattering see the text. \textit{Right:} range of plausible cooling times for synchrotron radiation (orange), bremsstrahlung (green), and the sum of all IC components (blue), resulting from a cluster photon field energy density of $U_\mathrm{cl} = 3\mbox{--}180\eVqcm$, a density inside the superbubble of $n_\mathrm{int} = 0.02\mbox{--}0.1\pqcm$, an effective cluster temperature of $T_\mathrm{eff} = 10{,}000\mbox{--}50{,}000\,$K, a magnetic field of $B = 1\mbox{--}10\mu$G, and an enhancement of the diffuse stellar and dust emission by a factor 1--3. The enhancement of the diffuse component is motivated by the proximity to the cluster and increased dust density in the region, compared to the standard ISM. The inset shows the behaviour in the TeV-band in more detail highlighting the loss-time scaling with energy. The default case (broad grey line) shows a behaviour that would require diffusion close to the Kraichnan regime ($D\sim E^{1/2}$) to reproduce energy-independent morphology.}
    \label{fig:tcool}
\end{figure*}

\begin{figure}
    \centering
    \includegraphics[height=0.25\textheight]{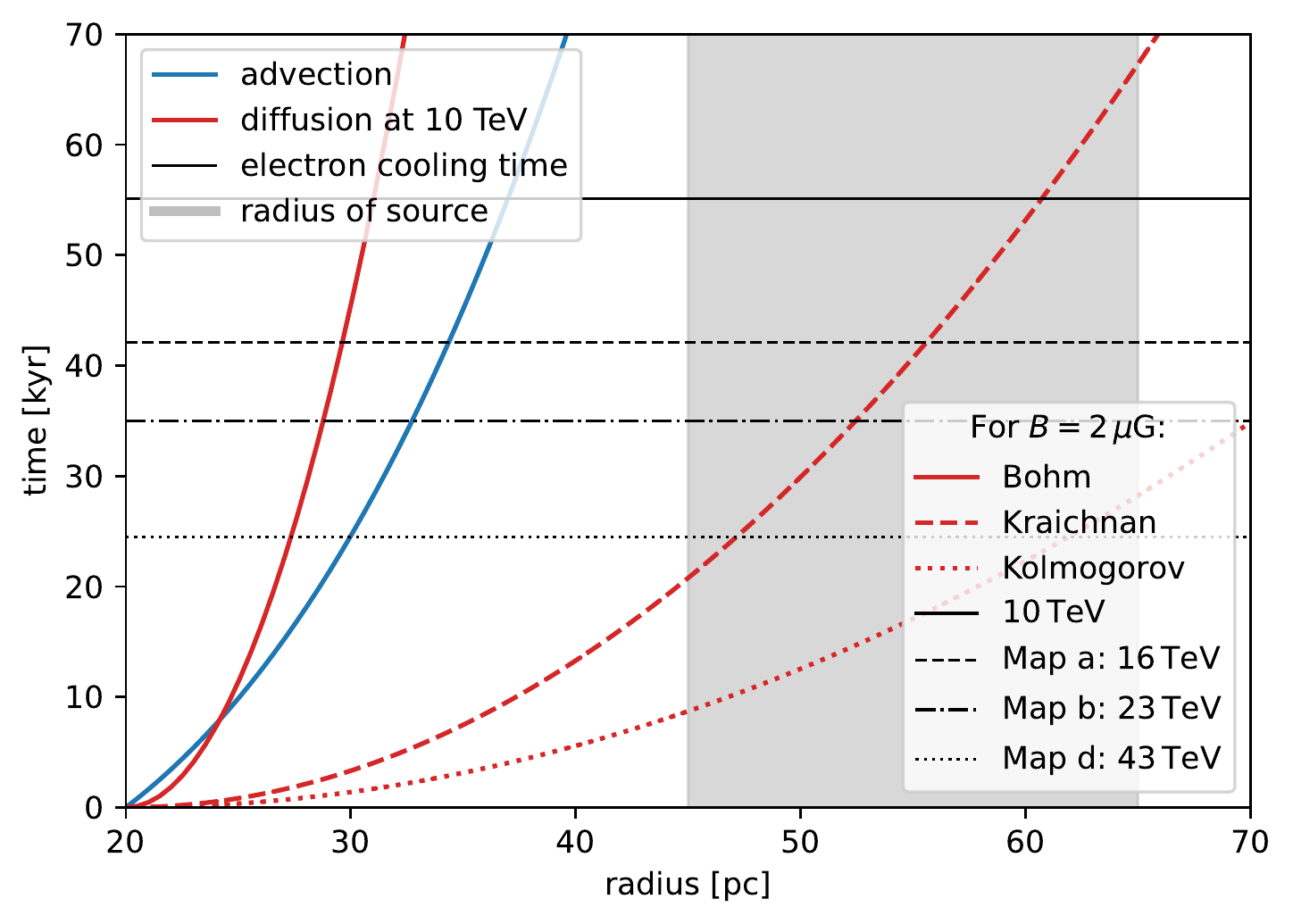}
    \caption{Diffusion (red) and advection (blue) timescales as a function of transported distance, assuming particles start at the termination shock at $R_\mathrm{ts} = 20.4\,$pc. Timescales for three different diffusion coefficients are shown, following Eq.~\ref{eq:bohmdiff} and \ref{eq:h_scaling}. The grey band indicates the radius of the source, where the large range being due to the elongated shape. Cooling times for electrons of four different energies are indicated (black). In addition to a fiducial value (10\,TeV), the average parent electron energies we determine for the \hess\ maps (see \citetalias{Aharonian22}, Fig.~3) are shown. The technique used is detailed in the text.}
    \label{fig:timescales}
\end{figure}

Electrons mainly cool by interacting with ambient photon fields and the magnetic field. We consider direct and dust-scattered starlight \citep[][]{Popescu17}, the Cosmic Microwave Background (CMB, 2.7\,K, $0.25 \eVqcm$) and the cluster photon field as discussed in Sect.~\ref{sec:clphotfield}. The model by \citet{Popescu17} is axisymmetric with respect to the GC and assumes a distance GC--Earth of 8\,kpc. We calculate the cooling times using the \texttt{GAMERA} library\footnote{\url{http://libgamera.github.io/GAMERA/docs/main_page.html}}, taking into account inverse Compton (IC) scattering on the above mentioned photon fields, synchrotron radiation, and bremsstrahlung. The results are shown in Fig.~\ref{fig:tcool}, on the left hand-side for a set of parameters that we adopt as default case in the following (see Table~\ref{tab:pars}) and on the right hand-side for limiting cases of maximal and minimal cooling (see caption of Fig.~\ref{fig:tcool}). Several conclusions can be drawn from the figure. First, the cooling break is below 100\,GeV, even in the case of minimal cooling. TeV particles are therefore expected to display a cooled spectrum. Second, cooling at 10--100\,TeV is dominated by IC scattering on the CMB and synchrotron radiation. Even though the cluster photon field has the highest energy density, it is strongly affected by the KN suppression in this energy range. The total cooling time at 10\,TeV is 55.1\,kyr for default parameters. The diffusion length is given by
\begin{equation}
    \ell_\mathrm{e} = 11 \left( \frac{D}{1.66\times 10^{26}\,\mathrm{cm}^2\,\mathrm{s}^{-1}} \right)^{0.5} \left( \frac{t_\mathrm{cool}}{55.1\,\mathrm{kyr}} \right)^{0.5} \mathrm{pc}\, ,
    \label{eq:ell_bl}
\end{equation}
where $D$ is the diffusion coefficient at the Bohm limit with $B = 2\,\mu$G. 
The magnetic field carried by the wind from the cluster is expected to be disordered on scales $\ll R_{\rm ts}$. Thus particles with gyro-radius equal to the outer-scale of the turbulence will have $h\approx1$, but will have an energy scaling that reflects the turbulence spectrum. 
Generally, $h$ scales as 
\begin{multline}
    h(E) = 
    \left( \frac{r_\mathrm{inj}}{1\,\mathrm{pc}} \right)^{1-\delta} \left( \frac{E}{10\,\mathrm{TeV}} \right)^{\delta - 1} \left( \frac{B}{2\,\mu\mathrm{G}} \right)^{1 - \delta}\\ \times
    \begin{cases}
        14 & \text{for Kraichnan, \,\,\,\,$\delta = 1/2$}\\
        33 & \text{for Kolmogorov, $\delta = 1/3$}
    \end{cases} 
    \, ,
    \label{eq:h_scaling}
\end{multline}
where $r_\mathrm{inj}$ is the injection scale of the turbulence, which can be assumed to be of the order of the average distance between stars in the cluster. The turbulence scaling inside SB is an open question, which is difficult to tackle with numerical simulations. Existing results are broadly consistent with $\delta \lesssim 0.5$ \citep[see][]{VieuPhD21}.

Figure \ref{fig:timescales} compares diffusion and advection times to the electron cooling time at different energies, chosen to predict the size of the source in the \hess\ maps with varying energy threshold (see Fig.~3 in \citetalias{Aharonian22}), which we discuss in detail in Sect.~\ref{sec:indep_morph}. The cooling time for 10\,TeV electrons is $\approx 70$\,kyr. In this time electrons are advected to a radius of ${\sim}37\,$pc approximately half way across the emission region. Diffusive transport, while uncertain, can account for the source size for range of values for $\delta$. The situation changes if the photon fields and the magnetic field deviate from our standard assumed values. With maximal IC cooling (see Fig.~\ref{fig:tcool}, right panel) the total cooling time at 10\,TeV is reduced to 28.7\,kyr, reducing the advection length by ${\sim}6\,$pc. This can be compensated if diffusion is closer to the Kolmogorov regime. In the case of larger magnetic field strength, both the cooling time and diffusion coefficient will reduce. For $B\gtrsim 5\,\mu$G, the transport should be diffusion dominated with $\delta \lesssim 1/2$. $B = 10\,\mu$G requires $\delta \approx 1/3$.

\subsection{Energy independent morphology}
\label{sec:indep_morph}
\citetalias{Aharonian22} show \hess\ maps for $E > 0.37$, ${>}1$, and ${>}4.9$\,TeV in their Fig.~3. Only a negligible change in source morphology is present between these bands. In this section, we discuss the implications of this energy independence for our model. For electrons, we predict the source size in each map based on the map's mean logarithmic $\gamma$-ray energy, where the mean is weighted with the log $\mathrm{d}N/\mathrm{d}E$ flux. Binned data from the combined energy spectrum are used (see Fig.~7 in \citetalias{Aharonian22}). The averages are 4.9, 8.3, and 20\,TeV for maps (a) $E > 0.37$, (b) ${>}1$, and (d) ${>}4.9$\,TeV, respectively. Next, we determine the cooling time of the parent electrons in the ambient photon fields. Electron energies are obtained by calculating IC spectra from several narrow-band electron injection spectra and selecting the spectrum with the highest flux at the given $\gamma$-ray energy, which yields 16, 27, and 43\,TeV for maps (a), (b), and (d), respectively. Figure \ref{fig:timescales} shows that the difference between map (a) and (d) is expected to be ${\sim}5\,$pc, which corresponds to ${\sim}4.4'$, for advection dominated transport, which is smaller than the kernel used to smooth the maps and therefore consistent with the observation of energy independent morphology. 

Energy independence is also achievable in the case of diffusion dominated transport. In the Thomson regime, $t_\mathrm{cool} \propto E^{-1}$, for Bohm diffusion, as $D\propto E$, the diffusion length is energy independent. In the transition to the KN regime, energy independent morphology can arise for other diffusion scenarios as well, for example Kolmogorov diffusion, where $D \propto E^{1/3}$, would have to be compensated by an $E^{-1/3}$ scaling of the cooling time. The inset in Fig.~\ref{fig:tcool} on the right hand-side illustrates this result. Energy independent morphology in the TeV arises in the Bohm case only if synchrotron cooling dominates and, in contrast, for Kolmogorov if IC cooling is strong and dominates. The default case requires diffusion close to Kraichnan scaling to reproduce energy independent morphology. However, due to the many uncertainties involved, no definite conclusions on the scaling should be drawn. 

Finally, consider the hadronic scenario. As the energy dependence of the proton cooling time is weak (see Eq.~\ref{eq:tcool_p}) and diffusion in the radial direction has to be suppressed, an energy independent morphology is trivially expected. Note, however, that the source morphology is hard to reconcile with a hadronic scenario and it is therefore disfavoured (see Sect.~\ref{sec:morph_protons}).  

\section{Spectrum}
\label{sec:spectrum}

\begin{figure}
    \centering
    \includegraphics[height=0.25\textheight]{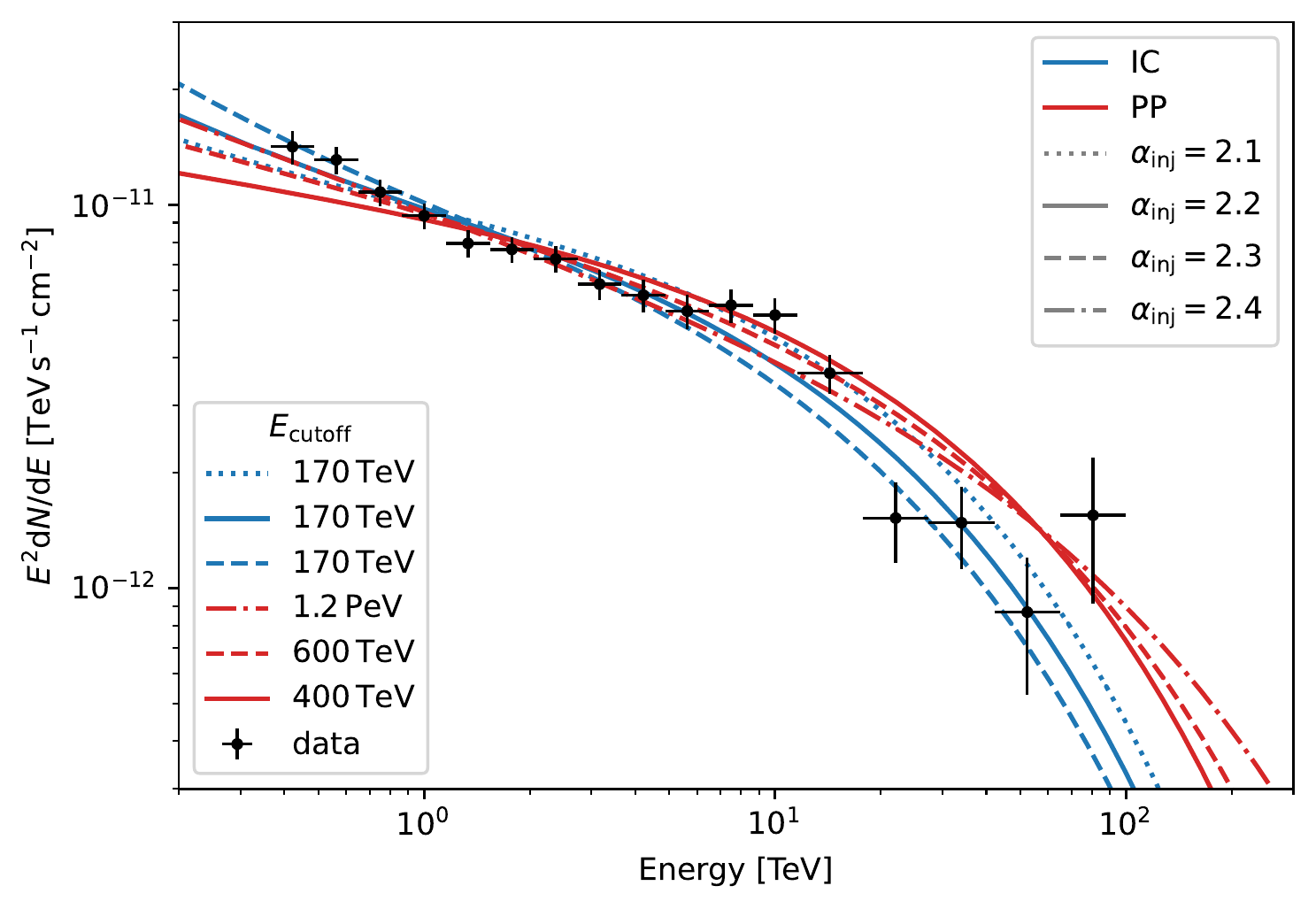} 
    \caption{The best models for the spectrum of HESS~J1646$-$458 (data taken from \citetalias{Aharonian22}). Standard parameters are assumed (see Table~\ref{tab:pars}). The injected particle spectrum is a power law with an exponential cut-off, where $\alpha_\mathrm{inj}$ is the index and $E_\mathrm{cutoff}$ the cut-off energy. The \texttt{GAMERA} library is used to calculate the cooled particle spectrum and $\gamma$-ray production.}
    \label{fig:hessband}
\end{figure}

We construct both a hadronic (PP interaction) and a leptonic (IC emission) model of the TeV $\gamma$-ray spectrum, using the \texttt{GAMERA} library to accurately include KN corrections. The particle injection spectra are set to be power laws with an exponential cut-off, $\text{d}N/(\text{d}E\text{d}t) \propto E^{-\alpha_\text{inj}} \exp(- E/E_\text{cutoff})$. The normalisation is obtained assuming that the power continuously injected into the particle spectra in the range 1\,GeV--1\,PeV is $\eta_\mathrm{p} L_\text{w}$ and $\eta_\mathrm{e} L_\text{w}$ for protons and electrons, respectively. In other words, the efficiencies $\eta_\mathrm{p}$ and $\eta_\mathrm{e}$ give the fraction of wind power converted into CR power at the termination shock. From the injection spectra, \texttt{GAMERA} is used to calculate the cooled spectra and consequentially the $\gamma$-ray emission. We use the GEANT 4.10.0 hadronic emission model and standard parameters for the leptonic emission. We consider the ambient radiation fields discussed in Sect.~\ref{sec:morph_e}: the cluster photon field, diffuse starlight, dust-scattered starlight, and the CMB. The best-fitting models are summarised in Fig.~\ref{fig:hessband}.

\subsection{Leptonic emission} 
\label{sec:em_lept}

\begin{figure}
    \centering
    \includegraphics[height=0.25\textheight]{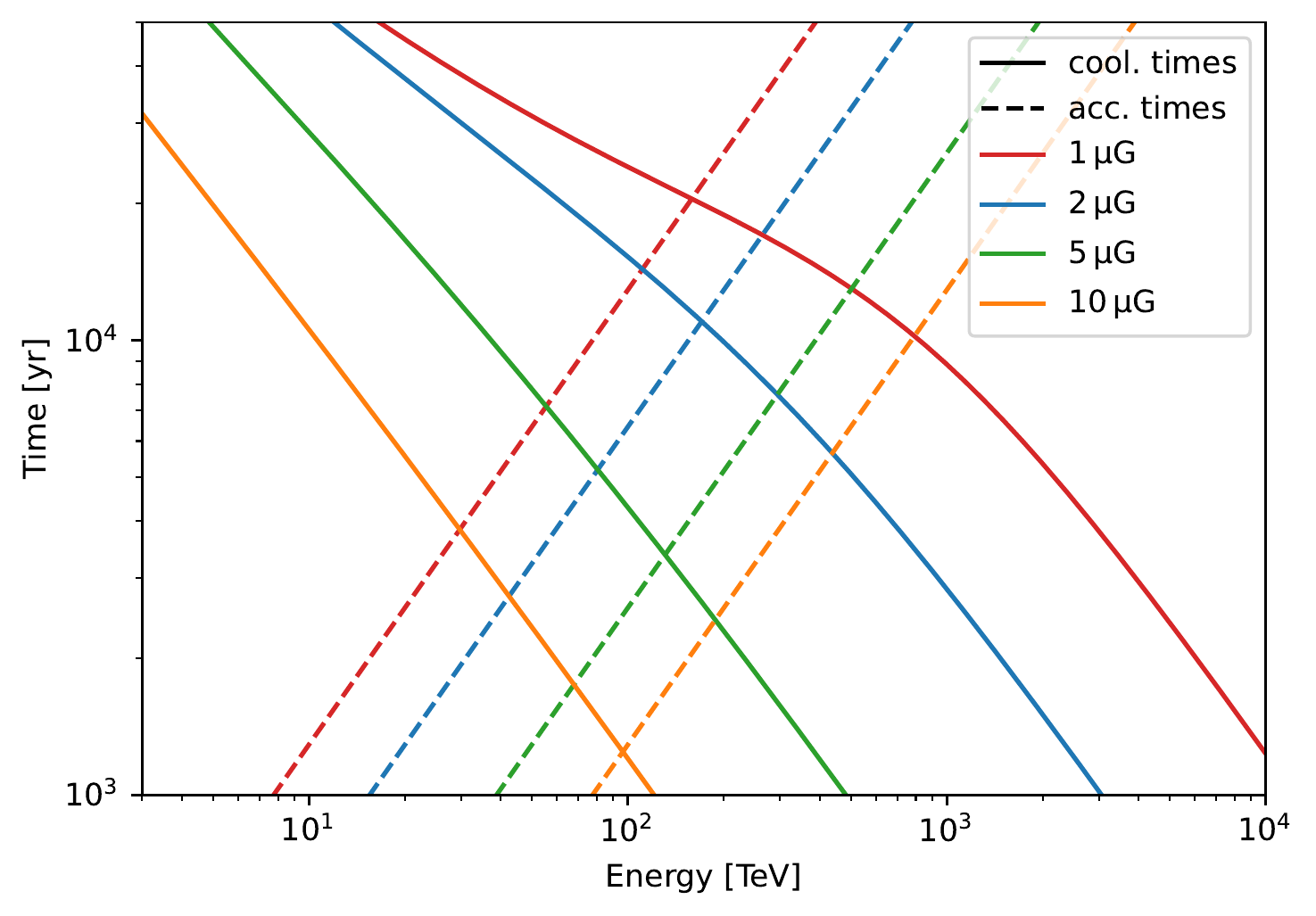}
    \caption{Cooling and acceleration times (solid and dashed lines, respectively) for a magnetic field in the acceleration region $B_\mathrm{acc} = 1\mbox{--}10\,\mu$G. The cut-off in the electron spectrum is to set the energy, where the two timescales are equal. Photon fields are chosen as in Fig.~\ref{fig:tcool}.}
    \label{fig:cutoffs}
\end{figure}

In the electron spectrum, $E_\mathrm{cutoff}$ is set to the energy where the cooling time equals the acceleration time, which assuming approximately equal upstream and downstream residence times is \citep{Drury83}
\begin{align}
    t_\mathrm{acc} &\approx \frac{8 D}{v_\mathrm{w}^2} = \frac{8h}{3} \frac{r_\mathrm{g}c}{v_\mathrm{w}^2}  
    \nonumber\\
    &= 6.7 h \left(\frac{E}{100\,\mathrm{TeV}}\right)\left(\frac{B_\mathrm{acc}}{2\,\mu\mathrm{G}}\right)^{-1}  \left(\frac{v_\mathrm{w}}{2500\,\mathrm{km~s}^{-1}}\right)^{-2} \,\mathrm{kyr} \, .
    \label{eq:tacc}
\end{align}
Figure~\ref{fig:cutoffs} shows the cooling and accelerations times for $B_\mathrm{acc} = 1\mbox{--}10\,\mu$G in the Bohm limit. Note that for $h > 1$, the acceleration time increases compared to Fig.~\ref{fig:cutoffs}, lowering $E_\text{cutoff}$. For the cooling, we consider the default case shown in Table~\ref{tab:pars} and in Fig.~\ref{fig:tcool} (left) and vary $B_\mathrm{acc}$. 
A $2\,\mu$G field places the cut-off in the particle spectrum at ${\sim}170$\,TeV. The resulting IC spectra (Fig.~\ref{fig:hessband}, blue lines) are consistent with the \hess\ data and require efficiencies of only $0.09\mbox{--}0.28\%$ for $\alpha_\mathrm{inj} = 2.1\mbox{--}2.3$ and default parameters (see Table~\ref{tab:pars}). For larger $B_\mathrm{acc}$, the IC model under-predicts the data at ${\gtrsim}10\,$TeV. We note that the magnetic field in the acceleration region $B_{\rm acc}$ need not be the same as the average value in the post-shock cooling region.

\begin{figure}
    \centering
    \includegraphics[height=0.25\textheight]{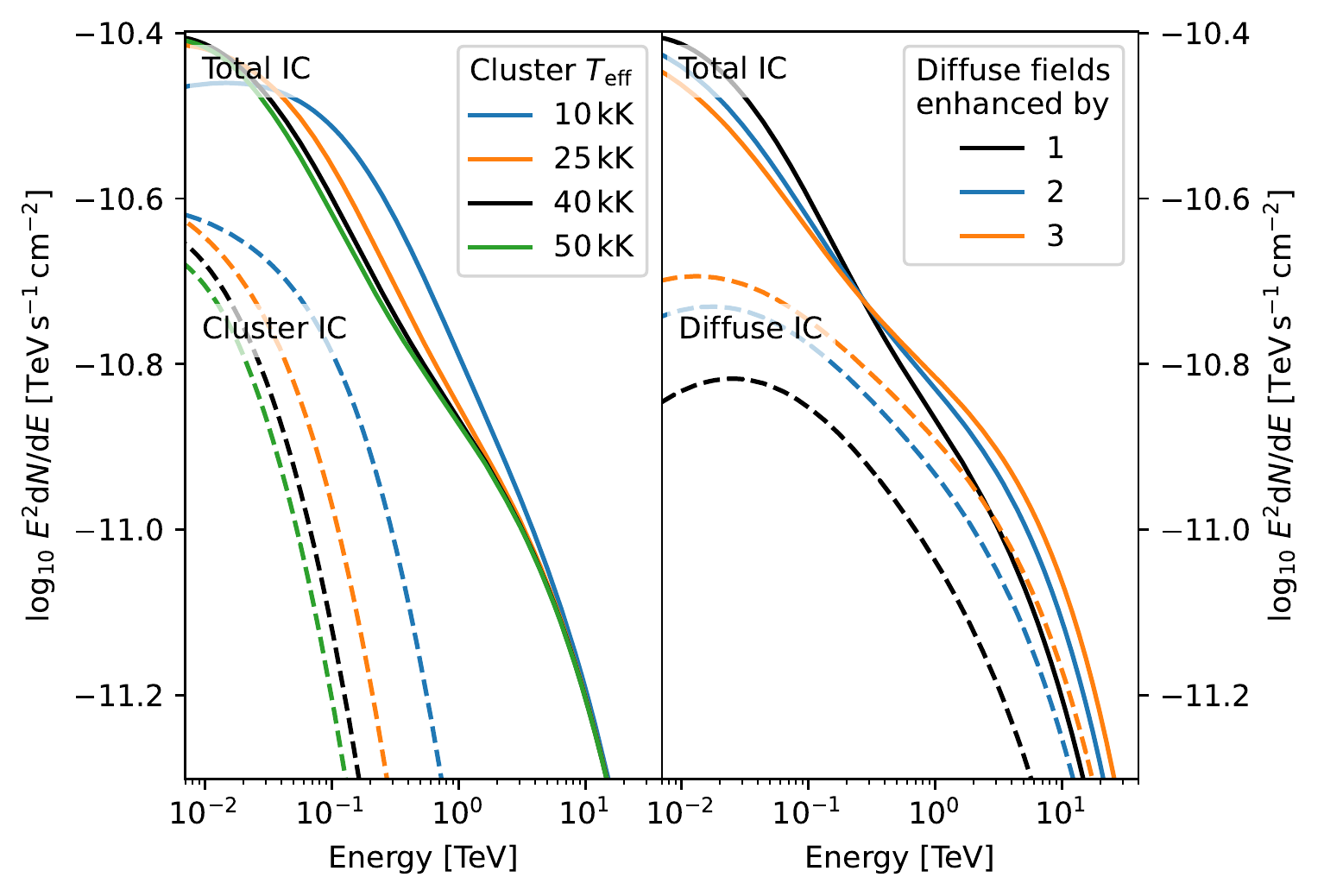}
    \caption{Effect of the variation of the cluster effective temperature ($T_\mathrm{eff}$, left) and an enhancement of the diffuse starlight and dust-scattered starlight radiation fields (right). Solid lines indicate the total IC spectrum. Dashed lines show only the cluster component on the left and the diffuse component on the right. Black lines show the best fitting result using default parameters (Table~\ref{tab:pars}), which is also displayed in Fig.~\ref{fig:hessband} with the blue dotted line. The normalisation of all models is set equal to the default case.}
    \label{fig:phfields}
\end{figure}

Figure~\ref{fig:phfields} illustrates the dependence of the IC spectrum on the photon fields. The left-hand panel concerns the cluster and diffuse photon fields. A low effective temperature steepens the spectrum between ${\sim}0.1\mbox{--}10\,$TeV and results in a higher flux, because the photons are less affected by KN suppression due to their lower energy. Such models are generally easier to reconcile with the slight steepening observed in the spectrum at ${\lesssim}1\,$TeV. For the diffuse fields (right-hand panel), we enhance the values predicted by \citet{Popescu17} by a factor 2--3. This accounts for reprocessed and lower energy cluster photons. An enhancement of the diffuse starlight and dust emission smooths the feature introduced by the cluster photon field at ${\sim}10\mbox{--}100\,$GeV, flattening the spectrum. In summary, IC emission explains the spectrum of HESS~J1646$-$458 well, although the required balance between cooling and acceleration times is a strong constraint on the model. 

The energetic electrons are also expected to produce synchrotron emission (see Fig.~\ref{fig:sync}). \citetalias{Aharonian22} deduced the brightness at 30\,GHz in a region with 1$^\circ$ radius around \src\ using data from the \textit{Planck} satellite, which can serve as an upper bound on the synchrotron component. Figure~\ref{fig:sync} shows that our models for $B = 1\mbox{--}10\,\mu$G are consistent with this bound.

\subsection{Hadronic emission}

Putting aside the above mentioned limitations on morphology and advection, we consider a single-zone model for hadronic emission. The spectral shape can be reproduced with $\gamma$-ray emission from PP interaction using $\alpha_\mathrm{inj} = 2.2\mbox{--}2.4$ and $E_\text{cutoff} = 400\,\mathrm{TeV}\mbox{--}1.2\,\mbox{PeV}$, although the required efficiency is an obstacle for the hadronic model. The standard setting, $L_\mathrm{w} = 10^{39}\ergs$ and $n_\mathrm{ext} = 100\pqcm$, requires $\eta_\mathrm{p}=26\mbox{--}92\%$. These values are optimistic, especially because we do not consider particle escape and the best-fit model corresponds to the upper bound of this range ($\alpha_\mathrm{inj} = 2.4$, $E_\text{cutoff} = 1.2$\,PeV). Harder injection spectra have trouble accounting for the steepening at ${\lesssim} 1 \,$TeV. The efficiency depends inversely on $L_\mathrm{w}$ and $n_\mathrm{int}$. An increase of their product by a factor of 4 brings $\eta_\mathrm{p}$ down to $6.5\mbox{--}23\%$, which is more plausible for shock acceleration. We noted in Sect.~\ref{sec:morph} that the size of the source is considerably over-predicted in the hadronic scenario unless $R_\mathrm{b} \lesssim 60\,$pc. From this constraint and the efficiency requirement set above, it follows that $n_\mathrm{ext} \gtrsim 300\pqcm$ (cf.~Eq.~\ref{eq:rts} and \ref{eq:rfs}), which is much higher than typical values (see Sect.~\ref{sec:bubble}). An alternative explanation is the presence of a spectral break not far below 1\,TeV. 

The above consideration concerns the case in which the $\gamma$-ray emission originates in the bubble interior. An estimate for the $\gamma$-ray luminosity expected from the SB shell, $L_\mathrm{sh}$, can be obtained from the ratio of the residence time in the shell, $t_\mathrm{res}$, to the proton cooling time (see Eq.~\ref{eq:tcool_p}). This yields
\begin{multline}
     L_\gamma^\mathrm{sh} \approx \eta_\mathrm{p} L_\mathrm{w} \frac{t_\mathrm{res}}{t_\mathrm{cool}} 
     = \eta_\mathrm{p} L_\mathrm{w} \frac{R_\mathrm{sh}^2}{4Dt_\mathrm{cool}} \\ 
     \hspace{1.3em} 
     = 3\times 10^{33} \frac{\eta_\mathrm{p} ({>}10\,\mathrm{TeV})}{0.01} \frac{L_\mathrm{w}}{10^{39}\,\mathrm{erg}\,\mathrm{s}^{-1}} \left(\frac{R_\mathrm{sh}}{10\,\mathrm{pc}}\right)^2 \\ \times  \left(\frac{D_\mathrm{sh} (10\,\mathrm{TeV})}{10^{29}\,\mathrm{cm}^2\,\mathrm{s}^{-1}}\right)^{-1} \frac{n_\mathrm{sh}}{200\pqcm} \, \mathrm{erg}\,\mathrm{s}^{-1} \, ,
     \label{equ:Lsh}
\end{multline}
where we have assumed that 1\% of $L_\mathrm{w}$ is converted to protons above 10\,TeV. This flux is more than an order of magnitude below that seen by \hess\ (see Sect.~\ref{sec:intro}). Since the shell emission is expected to be located at radii ${\gtrsim}60\,\mathrm{pc}$ and therefore is spread over a large solid angle, a non-detection in \hess\ is not surprising. Future observations may reveal the shell component, which would greatly enhance our understanding of particle acceleration in the region. Note, however, that the numerical values used in Eq.~\ref{equ:Lsh} are uncertain.

\section{Discussion and conclusion}
\label{sec:concl}

\begin{figure}
    \centering
    \includegraphics[height=0.27\textheight]{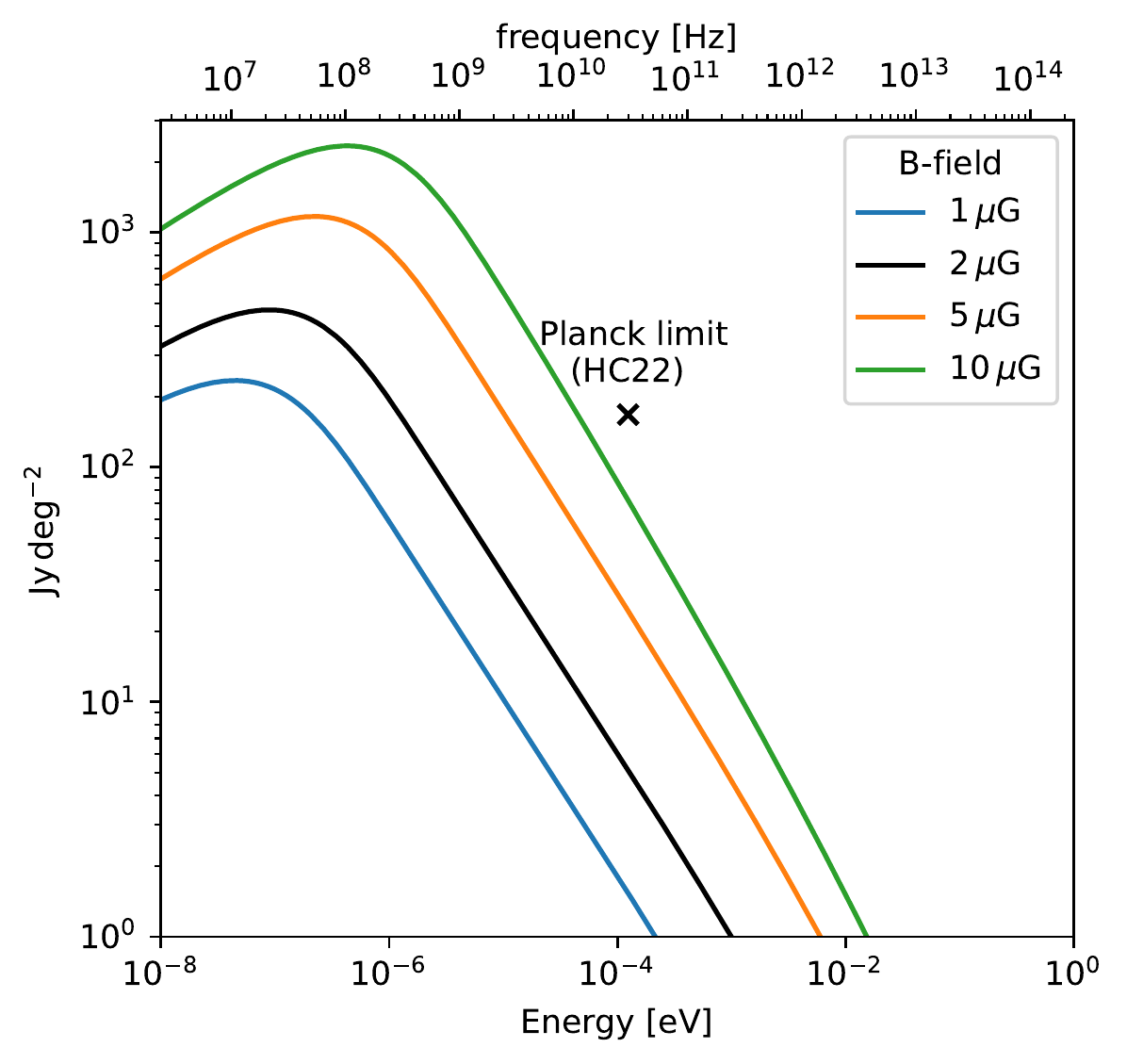}
    \caption{Predictions for the synchrotron brightness for an emission region magnetic field $B = 1\mbox{--}10\,\mu$G. The $2\,\mu$G model (black) uses the same set of parameters as the blue dotted model in Fig.~\ref{fig:hessband} and was used to set the normalisation for all models. For reference, the source covers ${\sim}1\,\mathrm{deg}^{2}$.}
    \label{fig:sync}
\end{figure}

We investigated the scenario of particle acceleration at a strong termination shock in a superbubble (SB) surrounding \src, with regard to the \hess\ observations reported in 2022 \citepalias{Aharonian22}. In Sect.~\ref{sec:morph} and \ref{sec:spectrum} we discussed how the TeV $\gamma$-ray morphology and spectrum of the source can be modelled with either inverse Compton (IC) emission or the decay of neutral pions produced via proton-proton (PP) interactions. 

The hadronic interpretation faces two main difficulties. Protons are expected to be uncooled for the given cluster age of 4\,Myr. Transport by both advection and diffusion over this timescale significantly overpredicts the size of the emission region. The only conceivable, though unrealistic, solution is a SB radius of ${\lesssim} 60\,$pc constraining the advection length and an essentially complete suppression of radial diffusion. The second issue is the energetics: $26\mbox{--}92\%$ of the cluster wind power have to be converted into $\gamma$-ray luminosity in the standard scenario. Alternative scenarios require unrealistic densities of the external medium of ${\gtrsim} 300\pqcm$, considering that morphology has to be accounted for as well. The efficiency issue is especially relevant for models with steep injection spectra, as they require more power at TeV energies than models with power law indices around 2. Figure~\ref{fig:hessband} shows that cut-off energies ${>}1\,$PeV require such steep injection spectra. A potential solution to the efficiency issue is a break not far below the \hess\ band. However, considering the general picture and that \citetalias{Aharonian22} did not find an alignment between the $\gamma$-ray emission and the neutral gas distribution, we deem a purely hadronic scenario, especially one where the particles reach PeV energies,  unlikely. 

The leptonic model provides an overall consistent picture that is in good agreement with the available data. We investigate IC scattering on diffuse starlight and dust-scattered starlight, the CMB, and a $T = 10{,}000\mbox{--}50{,}000\,$K thermal cluster photon field. Cooling shortens the advection and diffusion lengths considerably compared to the hadronic case. As a result, the model is consistent with both the radius of HESS~J1646$-$458 and the energy independent morphology. We note that the mean $\gamma$-ray energy of the \hess\ maps (Fig.~3 in \citetalias{Aharonian22}) only varies between 4.9 and 20\,TeV. The predicted energy dependence of the transport in this range is weak and not detectable given the resolution of the maps. Due to continuous injection by the star cluster and the Klein-Nishina (KN) suppression, which causes the cooling times of high energy electrons to rise, a sufficient supply of high energy electrons is present in the bubble, contrary to what was reported in \citet{Bhadra22}. In fact, we find that the efficiency required for the conversion of cluster wind power to IC luminosity is ${<}0.3\%$. The spectral shape is quite well predicted for an acceleration region magnetic field of $B_\mathrm{acc} \lesssim 2\,\mu$G. A higher $B_\mathrm{acc}$ lowers the cut-off energy and hence underpredicts the flux at ${\gtrsim}10\,$TeV. This constraint is the largest caveat of the leptonic model, but does not threaten its validity as $B_\mathrm{acc}\sim1\mbox{--}2\,\mu$G fulfils Hillas' limit and our requirement of a strong termination shock. If $B_\mathrm{acc} \gtrsim 2\,\mu$G, a hard hadronic component could compensate for the early cut-off in the electron spectrum. Such a joint model is less challenging than a purely hadronic model in terms of efficiency, but requires electrons and protons to be constrained within the same emission region which presents an additional assumption. The spectral shape of HESS~J1646$-$458 is well predicted by such models. Though the leptonic model is preferred globally, hadronic emission is expected at some level, and identifying its presence is a necessary step in determining the contribution of stellar clusters to the Galactic CR population.

A key open question is the position of the cut-off in the $\gamma$-ray spectrum, which southern hemisphere $\gamma$-ray observatories such as CTA and SWGO will be able to address. Radio data could constrain the magnetic field in the emission region, improving upon the upper bound set by \citetalias{Aharonian22} (see Fig.~\ref{fig:sync}). The GeV-band can also provide valuable insight, as many of the models shown in Fig.~\ref{fig:hessband} separate in this range. The IC emission from the cluster photon field also peaks in the GeV (see Fig.~\ref{fig:phfields}) and is key in understanding the spectral behaviour. 

\begin{acknowledgements}
We thank Giada Peron, Richard Tuffs, and Felix Aharonian for helpful discussions. We also thank the referee for their comments, which helped improving the manuscript.
\end{acknowledgements}


\bibliography{main}
\bibliographystyle{aa}

%\begin{thebibliography}{99}
%\end{thebibliography}

%-------------------------------------------------------------
%                   For appendices and landscape, large table:
%                    in the preamble, use: \usepackage{lscape}
%-------------------------------------------------------------

% \begin{appendix}
% The appendix

% \end{appendix}
\end{document}